\documentclass[prb,aps,reprint]{revtex4-1}

\usepackage{graphicx}
\usepackage{amsmath}
\usepackage{amssymb}
\usepackage{mathtools}

\begin{document}


\title{Low temperature magnetoresistance of (111) (La$_{0.3}$Sr$_{0.7}$)(Al$_{0.65}$Ta$_{0.35}$)/SrTiO$_3$}

\author{V. V. Bal$^1$, Z. Huang$^{2,3}$, K. Han$^{2,3}$, Ariando$^{2,3}$, T. Venkatesan$^{2,3,4,5,6}$, \& V. Chandrasekhar$^{*,1}$}
\affiliation{$^1$Department of Physics and Astronomy, Northwestern University, Evanston, IL 60208, USA.\\ 
 $^2$NUSNNI-Nanocore, National University of Singapore, 117411 Singapore.\\
 $^3$Department of Physics, National University of Singapore, 117551 Singapore.\\
$^4$NUS Graduate School for Integrative Sciences and Engineering, National University of Singapore, 117456 Singapore.\\
$^5$Department of Electrical and Computer Engineering, National University of Singapore, 117576 Singapore.\\
$^6$Department of Material Science and Engineering, National University of Singapore, 117575 Singapore.}

\date{\today}                              
                            
\begin{abstract}
 The two dimensional conducting interfaces in SrTiO$_3$-based systems are known to show a variety of coexisting and competing phenomena in a complex phase space. Magnetoresistance measurements, which are typically used to extract information about the various interactions in these systems, must be interpreted with care, since multiple interactions can contribute to the resistivity in a given range of magnetic field and temperature. Here we review all the phenomena that can contribute to transport in SrTiO$_3$-based conducting interfaces at low temperatures, and discuss possible ways to distinguish between various phenomena. We apply this analysis to the magnetoresistance data of (111) oriented (La$_{0.3}$Sr$_{0.7}$)(Al$_{0.65}$Ta$_{0.35}$)/STO (LSAT/STO) heterostructures in perpendicular field, and find an excess negative magnetoresistance contribution which cannot be explained by weak localization alone.  We argue that contributions from magnetic scattering as well as electron-electron interactions can provide a possible explanation for the observed magnetoresistance.
\end{abstract}

\maketitle
\section{Introduction}
Since its discovery in 2004,\cite{Ohtomo} the two dimensional conducting gas (2DCG) at SrTiO$_3$ (STO)-based complex oxide interfaces has proven to be a fertile ground for the study of a great variety of physical phenomena.\cite{Brinkman,Reyren,Thiel,Caviglia,Dikin,Bi,Levy,Ariando} The electronic structure of these systems is characterized by the presence of multiple, anisotropic bands at the Fermi surface,\cite{Santander,Joshua,Syro} multivalent transition metal ions, a high degree of electronic correlations, and the breaking of inversion symmetry.  This structure can be modified due to the ease of doping with oxygen vacancies\cite{Siemons,Ariando2,Davis} as well as other cations, the propensity to electronic and structural reconstructions and phase transitions, and strain.\cite{Eom,Huang} Additionally, the high dielectric constant of STO which can be tuned by an electric gate voltage, $V_g$, allows for an \textit{in-situ} modulation of sample properties.\cite{Hemberger} All these factors make for a complicated phase space, with phenomena including superconductivity,\cite{Reyren,Thiel,Han,Monteiro} superconductor-insulator transitions,\cite{Mehta,Mehta2} charge ordering,\cite{Sam} and magnetic behavior.\cite{Brinkman,Dikin,Ashoori,Moler,Bi,Ariando}

An important goal is to understand what interactions within STO-based 2DCGs lead to these varied behaviors, and how we can tune a particular physical parameter to control the interactions. Magnetoresistance (MR) studies in fields perpendicular and parallel to the 2DCG, in conjunction with temperature dependence measurements, are often used to shed light on the band structure and unravel the different mechanisms in the system, which typically show different dependencies on magnetic field scale, field orientation, and temperature. However, the situation in the case of STO-based 2DCGs is not straightforward, owing to the many degrees of freedom this electronic system possesses. The mobility and density of the multiple types of carriers present at the Fermi surface, which originate from the interfacial Ti 3$d$ $t_{2g}$ orbitals of STO,\cite{Santander,Walker} can be tuned by $V_g$, and the magnetic interactions between localized moments and/or itinerant carriers can be modified as a result.\cite{Bi,Ruhman,Bal} The presence of strong electron-electron interactions (EEI),\cite{Levy,Nandy,Fuchs} superconductivity,\cite{Reyren,Thiel} as well as spin-orbit interactions (SOI)\cite{Caviglia} is also controlled by $V_g$. Finally, the inherent disorder in the system, which gives rise to localization,\cite{Hernandez,Caviglia} is also dependent on $V_g$. All these phenomena contribute to sample resistivity at low temperatures, and must be accounted for when trying to understand transport in this system. 

So far, research efforts have mainly focused on the (001) oriented STO-based 2DCGs. However, the (110) and (111) oriented heterostructures have recently been shown to host 2DCGs with very interesting properties.\cite{Herranz,Davis2,Davis,Dagan,davis3,davis4} In this paper, we analyze the MR data in perpendicular fields $B$ for the 2DCG in (111) oriented (La$_{0.3}$Sr$_{0.7}$)(Al$_{0.65}$Ta$_{0.35}$)/STO (LSAT/STO) heterostructures. This particular system is interesting owing to the hexagonal symmetry of the Ti 3$d$ $t_{2g}$ orbitals, and has been predicted to show topological physics.\cite{Syro,Walker,Pickett,Xiao} LSAT also has a smaller lattice mismatch with STO in comparison to LaAlO$_3$ (LAO). This gives rise to a smaller strain in the LSAT/STO system as compared to the more widely studied LAO/STO system, which can lead to higher carrier mobilities,\cite{Huang} as well as modify the orbital ordering of the LSAT/STO 2DCG.\cite{Pickett,Xiao} 

We have previously shown qualitatively that the SOI in the (111) LSAT/STO 2DCG increases as $V_g$ is reduced,\cite{Bal2} in contrast with what has been observed in case of (001) STO-based 2DCGs,\cite{Caviglia} and that at millikelvin temperatures, ferromagnetic order, characterized by hysteresis in the MR, emerges as the SOI becomes stronger at low values of $V_g$.\cite{Bal} In this paper, we discuss the quantitative analysis of the MR in STO-based 2DCGs, and in (111) LSAT/STO in particular. We argue that obtaining quantitative values of the phase coherence length $l_\phi$ and the spin-orbit scattering length $l_{so}$ is complicated by the possible presence of magnetic scattering and EEI which result in an excess negative MR. 

The rest of the paper is organized as follows: In Section II we review the various mechanisms that contribute to the resistivity of STO-based 2DCGs, along with their field and temperature dependencies. In Section III we describe our sample fabrication and measurement methods, and in Section IV we present the analysis of our MR data on (111) LSAT/STO. We show that we can fit our data up to $B \sim$ 3 T in terms of weak localization/antilocalization corrections, by accounting for a background term which is second order in $B$. This background term comes from a combination of a positive MR due to the classical orbital contribution, a negative MR that is quadratic at low fields and saturates at high fields, likely caused by magnetic scattering, and EEI effects which can be positive or negative. We quantitatively demonstrate that SOI in the (111) LSAT/STO 2DCG increases, and the phase coherence length decreases, with decreasing $V_g$.

\section{Contributions to resistance}
For STO-based 2DCGs, the sheet resistance $R$ which depends on carrier density $n$ and mobility $\mu$ as $R = 1/ne\mu$, can change by orders of magnitude when $V_g$ is changed from large positive values to negative values (typically a few tens of volts to hundreds of volts, both positive and negative). In contrast, the Hall coefficient $R_H = 1/ne$ typically changes only by less than a factor of 2 or 3.\cite{Triscone, Hurand, Davis} This suggests that the change in resistance as a function of $V_g$ is a result of a large change in carrier mobilities, which depend on scattering time $\tau$ and effective mass $m^*$ as $\mu = e\tau/m^*$, rather than a change in carrier densities. This trend in $R$ and $R_H$, which is a common feature of STO-based 2DCGs,\cite{Dikin,Caviglia,Davis} is also observed in our sample \cite{Bal2} and warrants further investigation to understand the causes of the drastic change in $\mu$, or equivalently, in $\tau$. 

For STO-based 2DCGs in general, the sheet resistance is known to show a minimum at a temperature of a few Kelvin, increasing in value as temperature is lowered further, before finally either saturating, or vanishing if the sample undergoes a superconducting transition \cite{Brinkman,Reyren,Huang,Bal,Goldhaber} depending on growth conditions and the particular value of $V_g$. Hence the MR at sub-Kelvin temperatures for different values of $V_g$ can give us important information about the scattering mechanisms that lead to the aforementioned drastic changes in $R$ as a function of $V_g$, given that these changes are amplified at lower values of $T$.

Various scattering processes exist in a system, and are modulated by factors such as $T$, $V_g$, and $B$. We now look at the contributions to $R$ due to each of these processes, and discuss, in the context of STO-based 2DCGs, how they affect $R(T,B)$ as the disorder, dimensionality, SOI, and the multiband nature of the system is changed.



\subsection{Magnetic Field Independent Contributions}
\noindent{\textit{Drude contribution} ($R_0$): In metallic systems, the sheet resistance at zero field, $R_0$, independent of $T$ and $B$, is the Drude contribution, caused by the elastic scattering of carriers off static impurities and surfaces, and can be calculated in terms of the transport scattering time ($\tau$), carrier density ($n$), and carrier mass ($m^*$). In 2D systems, in which conductivity is the same as conductance, the Drude contribution can be written as $R_0 = m^*/ne^2\tau$, where $n$ is the areal charge density. 
If contributions due to other mechanisms are small compared to $R$, then the resistance $R$ can be approximated as $R_0$ for the purpose of determining $\tau$. However, in a real system where other contributions are substantial and difficult to pry apart, it is unclear that the measurement of $R$ at any given temperature gives us the value of $R_0$. In the case of STO-based 2DCGs, this is especially a problem in case of measurements at negative values of $V_g$, for which resistance changes rapidly as a function of $T$ at the low temperatures of interest, and the Drude picture may not apply. For our (111) LSAT/STO sample, this can be seen clearly from Fig. 1, where for $V_g$ = -40 V, $R$ changes by over 15\% between $T$ = 500 mK and $T$ = 50 mK. The situation is more amenable to analysis for the range of $V_g$ studied in more detail here, i.e., $V_g \geq$ 60 V, where $R$ changes by less than 5\% over the temperature range of interest. 

The effective carrier mass $m^*$ is typically obtained from angle resolved photo-emission spectroscopy (ARPES) measurements. ARPES studies on vacuum cleaved (111) STO have revealed highly anisotropic effective masses of electrons from the Ti 3$d$ $t_{2g}$ orbitals of interest, with a heavy (light) mass of 1.8 $m_e$ (0.27 $m_e$) along the [1$\bar{1}$0] direction, and a heavy (light) mass of 8.67 $m_e$ (0.33 $m_e$) along the [$\bar{1}\bar{1}$2] direction, with $m_e$ being the bare electron mass.\cite{Syro} An estimate for the cyclotron $m^*$ can also be obtained from an analysis of Shubnikov-de Haas (SdH) data. In the case of our (111) LSAT/STO sample, we do not see enough SdH oscillations within the range of field available to us (10 T) to get a reliable estimate of $m^*$.\cite{Bal2} For rough estimates of various parameters, we have used $m^* = m_e$ for simplicity, however this does not in any way affect the main results of our analysis. 

The carrier density $n$ is estimated using Hall data. Hall data in STO-based interfaces are electron-like, and show nonlinear behavior, especially at higher values of $V_g$. This has been interpreted as evidence of multicarrier transport. At lower values of $V_g$ hole-like carriers are also believed to play a role.\cite{Joshua,Davis,Bal} Hence the estimate of $n$ obtained from Hall measurements may be a good approximation only when single band transport dominates, possibly in the case of low densities or low values of $V_g$, but not if multiple bands are involved in transport. This in turn introduces uncertainty in the straightforward determination of $\tau$, which complicates the determination of other transport parameters, namely, the Fermi wavenumber $k_F = \sqrt{2 \pi n}$, Fermi velocity $v_F = \hbar k_F/m^*$, mean free path $l$, and the diffusion constant $D = v_F l/2$ for two dimensional systems.\\}

\noindent{\textit{Contribution due to phonon scattering} ($\Delta R_{ph} (T)$): Inelastic scattering of electrons off phonons leads to a contribution with a power law temperature dependence. The electron-phonon contribution is proportional to $T^5$ in the clean limit in the case of simple isotropic metals, or proportional to $T^3$ if Umklapp scattering is dominant.\cite{Santhanam} 
Other powers are also possible if multiple types of scattering mechanisms are present.\cite{Wiser} We note that these contributions are not expected to play a role in the temperature range under study in this paper, since these scattering mechanisms are frozen out to a large extent at very low temperatures. In STO-based systems, many experiments have identified a $T^2$ dependence of $R$,\cite{Fuchs,Behnia, Baber} attributed to phonon-mediated electron-electron scattering, or electron-electron scattering in the presence of multiple bands.\\}

\noindent{\textit{Contribution due to charged impurities} ($\Delta R_{ion} (T)$): Charged impurities such as oxygen vacancies are a common occurrence in STO. These occur, for example, when the system is annealed in a reducing atmosphere, causing the removal of neutral oxygen atoms from the crystal. This leaves behind two extra electrons in the crystal. Near the interface, these oxygen vacancies form a donor level just below the conduction band (which is composed of 3$d$ orbitals) of the system. The extra electrons can be excited into the conduction band if the temperature is high enough, and participate in transport. However, as $T$ is reduced, electrons can drop back into the donor level, in effect being trapped by the positively charged oxygen vacancy sites. These charge traps are known to have activation temperatures $T_A$ ranging from a few Kelvin to a few tens of Kelvins.\cite{Ariando2} The concentration of these charged impurities decreases exponentially with increasing temperature on the scale of $T_A$. Also, the screening of these impurities decreases with increasing temperature, since the dielectric permittivity of STO, which is also a function of $V_g$, decreases with increasing temperature.\cite{Barrett} Scattering of electrons off these partially screened charged impurities leads to the contribution $\Delta R_{ion} (T)$, which when combined with the change in resistivity caused by the inelastic mechanisms described in the previous paragraph, can lead to a resistance minimum at intermediate temperatures, with low temperature saturation, that is commonly observed in STO based 2DCGs.\cite{Fuchs} This mechanism may be present in combination with the Kondo mechanism, which is typically used in order to describe the observed resistance minimum in these systems,\cite{Brinkman,Goldhaber} and which will be discussed later.\\}

\subsection{Magnetic Field Dependent Contributions}
\noindent{\textit{Classical orbital contribution} ($\Delta R_{cl} (T,B)$): A magnetic field perpendicular to the 2DCG causes an increase in path length and back-scattering of electrons due to orbital effects. If only electron-like (or hole-like) carriers from closed bands participate in transport in a clean system (with one dominant carrier mobility), $\Delta R_{cl} (T,B)$ is proportional to $B^2 \sim (\omega_c \tau)^2$ for low fields ($\omega_c \tau <1$), where $\omega_c = eB/m^*$ is the cyclotron frequency, while the MR saturates at high fields ($\omega_c \tau > 1$). In the case of STO-based systems, a quasilinear behavior is typically observed at high fields,\cite{Dagan2,Bal} indicative of some degree of hole transport, or disorder (large spread in carrier mobility) in the 2DCG.\cite{Parish} In high mobility STO-based 2DCGs,\cite{Dagan2} for large positive values of $V_g$ where multiple bands contribute to transport, this $\Delta R_{cl} (B)$ can be very large, comparable to any low field corrections to the MR at fields as small as a few 100 mT,\cite{Bal2} and must be taken into consideration as a background while analyzing the low field MR. As the scattering time $\tau$ increases with decreasing temperature, this contribution increases with decreasing $T$.\\}

\noindent{\textit{Contributions due to magnetic scattering} ($\Delta R_{mag} (T,B)$): Going from higher to lower values of $V_g$, the size of $\Delta R_{cl} (B)$ is observed to reduce considerably, and in some cases a negative MR emerges at the lowest values of $V_g$.\cite{Joshua,Bal} This MR is seen to remain negative even at the highest values of $B$ studied. One of the causes of negative MR is the presence of magnetic scattering in the system. STO-based 2DCGs are known to show a wide range of magnetic phenomena, ranging from Kondo-like behavior caused by dilute magnetic scatterers in the system, to spin glasses, to a full ferromagnetic phase at the highest concentration of magnetic scatterers.\cite{Ruhman} What is more, these three regimes may coexist in the 2DCG owing to a disordered distribution of magnetic scatterers. For all these regimes, however, a negative isotropic MR has been predicted and observed in many systems including STO-based 2DCGs.\cite{Nigam, Inaba,Brinkman} This negative contribution to the MR, $\Delta R_{mag} (T,B)$, which can be large, is proportional to $B^2$ for smaller fields, and saturates at higher fields greater than those required to saturate the magnetic moments. This negative MR must also be considered on a similar footing as the positive $\Delta R_{cl} (B)$ in order to analyze the low field corrections.}

The temperature dependence of resistivity due to the presence of magnetic scatterers depends on the whether the magnetic moments are in the dilute or the spin glass limit. In both situations the resistivity increases logarithmically as temperature is decreased, as conduction electrons scatter off partially screened magnetic moments. If the temperature is lowered below the characteristic Kondo temperature of the system, the moments are fully screened, and a saturation in resistance is observed. If, however, the concentration of magnetic moments is high, and if the temperature is low enough that the thermal energy is smaller than the strength of interaction between individual magnetic moments, the moments start to freeze out, leading to a spin-glass phase, wherein the sample resistance can even decrease as temperature is lowered.\cite{Jonghwa} Thus the presence of magnetic moments in STO-based 2DCGs can be invoked to explain some of the observed $T$ and $B$ dependencies in this system. 

It was discussed earlier that scattering of conduction electrons off ionic impurities can give a similar temperature dependence as the scattering of conduction electrons off magnetic moments. In principle, it should be possible to tell these two mechanisms apart by measuring the temperature dependence of resistance while applying a magnetic field. $\Delta R_{ion} (T)$ should remain unaffected by $B$, while in the case of $\Delta R_{mag} (T, B)$, the resistance minimum and low temperature saturation should progressively disappear for higher values of $B$. However, the presence of localization corrections and EEI corrections (to be discussed later) also can give rise to a difference in the temperature dependence of resistivity for different values of $B$. Another way would be to look for a peak in specific heat of the sample near the estimated Kondo temperature, however, to our knowledge, this technique has not been used so far in the case of STO-based 2DCGs.\\

\noindent{\textit{Single particle localization contributions} ($\Delta R_{loc} (T,B)$): In two dimensions in the presence of disorder, and in the absence of SOI, all electronic states are localized at zero temperature.\cite{Anderson} If disorder is strong, i.e., $k_Fl <1$ or equivalently, $R > R_Q = $ 25.812 k$\Omega / \Box$, which is the quantum of resistance, then $\Delta R_{loc} (T,B)$ increases exponentially as a function of $T$.\cite{Ramakrishnan} For our sample, even at the lowest value of $V_g$ studied, $R$ at $T$ = 50 mK is $\sim$ 33 k$\Omega / \Box$, only marginally greater than $R_Q$. 

In the regime of $R$ for our sample, the predictions of the weak localization theory, which assumes a diffusive system and employs perturbative techniques to derive single-particle corrections to the conductivity resulting from the constructive interference of coherently back-scattered carriers, are generally valid.\cite{Ketterson} In two dimensions, weak localization predicts a logarithmic increase in resistance as $T$ is reduced  as given by Eqn. 1.\cite{Ramakrishnan}}

\begin{equation}
\Delta R_{loc} (T,0) = - \frac{R_0^2}{2 \pi^2 \hbar/e^2} \hspace{1mm}p       
\hspace{1mm} \mathrm{ln}\frac{T}{T_0} 
\end{equation}
where $T_0 = \hbar/k_B \tau$. The effect is caused by an increase in the phase coherence time $\tau_\phi$ with decreasing temperature, which typically goes as $T^{-p}$,\cite{Ramakrishnan} where $p$ depends on the mechanism of decoherence. An applied magnetic field perpendicular to the 2DCG also impedes the coherent interference of the backscattered electron waves, and leads to a MR. The sign and magnitude of this MR depends on not just $\tau_{\phi}$, but also on $\tau_{so}$, the spin-orbit scattering time, and $\tau_s$, the spin-flip scattering time. The form is also dependent on the type of SOI present in the system, i.e., whether it has a cubic or a linear dependence on momentum.\cite{Punnoose,Nayak} If SOI is substantial, then Zeeman effects can play a role as well, with the electron $g$ factor of the 2DCG as an additional parameter. \cite{Maekawa} Finally, the exact form of the $T$ and $B$ dependencies are dictated by the dimensionality of the system with respect to weak localization, i.e., if the associated length scale for decoherence, $l_\phi = \sqrt{D \tau_\phi} $ is greater than the film thickness $d$, then the film is in the two-dimensional limit. 

Despite the complexity of the various theories, it is clear that SOI is an antilocalizing mechanism, since the spin-rotation caused by SOI leads to an increase in the destructive interference of coherently backscattered carriers. Hence an applied $B$, which causes decoherence, causes a negative MR in the absence of strong SOI, and a positive MR in the presence of a strong SOI. The role of magnetic scattering is also to cause decoherence.\cite{Bergmann} The changes in conductivity due to weak localization/antilocalization are of the order of $\sigma_0 = 2e^2/h$, while the field scales of the effects, $B_\alpha$, depend on $D$ and the relevant scattering time $\tau_\alpha$ as $B_\alpha = \hbar / 4eD\tau_\alpha$. Estimates of $B_\alpha$ are obtained by fitting to Eqn. 2 which describes $\Delta R_{loc} (T,B)$:\cite{Hikami} 

\begin{multline}
\frac{\Delta R_{loc} (T,B)}{R_0} = \frac{R_0}{2 \pi^2 \hbar/e^2} \Big[-\frac{3}{2} \Psi \Big( \frac{1}{2} + \frac{B_2}{B} \Big) + \\
\frac{1}{2} \Psi \Big( \frac{1}{2} + \frac{B_1}{B} \Big) + \mathrm{ln}\frac{B_0}{B}\Big].
\end{multline}

Here $B_1 = B_\phi + 2 B_s$, while $B_2 = B_\phi + (4/3) B_{so} + (2/3) B_s$, and $B_0$ is the field associated with the elastic scattering time $\tau$.
When analyzing the normalized differential MR data, the elastic field $B_0$ drops out of the equation, as we shall discuss in Section IV. In the context of STO-based 2DCGs, it is difficult to obtain reliable estimates of the diffusion constant $D$ of the system as discussed earlier, hence we describe $\Delta R_{loc} (T,B)$ in terms of characteristic length scales $l_\alpha$ instead of $\tau_\alpha$, with $l_\alpha^2 = \hbar/4eB_\alpha$. The temperature dependence of magnetic scattering in this system is also unknown. To minimize the number of fit parameters in the analysis, we ignore $B_s$, which is the contribution of magnetic scattering, as well as the Zeeman effect to weak localization corrections, and use the form in Eqn. 2 derived by Hikami, Larkin and Nagaoka, which considers the effect of SOI only as a scattering rate, without using the forms specific for linear or cubic SOI. 

The corrections due to localization are expected to decrease with increasing $B$, and completely die out at $B \sim \hbar/l^2e$.\cite{Bergmann} \\


\noindent{\textit{Contributions due to EEI in the diffusive limit} ($\Delta R_{EEI} (T,B)$): EEI effects contribute to the sample resistance in a number of ways. Large angle inelastic collisions in the ballistic limit contribute to the $T^2$ dependence of $\Delta R_{in} (T)$ discussed earlier. Small as well as large angle collisions can modify single particle lifetimes of electrons and cause the decoherence of electron wavefunctions, thus affecting $\tau_\phi$ which in turn affects the localization corrections. On the other hand, many-body EEI effects in the diffusive limit  can cause a change in the density of states of the 2DCG, and lead to the following corrections to the conductivity:\cite{Altshuler}}

\begin{widetext}
\begin{equation}
\Delta \sigma_{EEI} (T,B)  =  \frac{e^2}{\hbar} \frac{1}{4 \pi^2} \Big( 2 - \frac{3F}{2} \Big) \mathrm{ln} \Big( \frac{k_B T \tau}{\hbar} \Big) \\
-  \frac{e^2}{\hbar} \frac{1}{4 \pi^2} F g_2\Big( \frac{g\mu_BB}{k_BT}\Big) \\
-  \frac{e^2}{\hbar} \frac{1}{4 \pi^2} g_1(T) \Phi_2 \Big( \frac{2DeB}{\pi k_BT} \Big)
\end{equation}
\end{widetext}

Here the first term is the field-independent exchange and singlet Hartree contribution of the particle-hole channel, the second term is the triplet Hartree contribution, while the third term is the orbital contribution due to the particle-particle channel.\cite{Ramakrishnan} $F$ and $g_1 (T)$ are both related to the screened Coulomb potential. Since typically $|g_1 (T)| << 1$, this term is usually ignored. $F$ is of the order of unity and hence the first two terms of the equation must be considered in our analysis. 

The second term gives a negative correction to the conductivity, and hence a positive $\Delta R_{EEI} (B)$. $g_2 (T,B)$ has a functional form, $\sim 0.084 (g\mu_B B/k_BT)^2$ for $g\mu_B B/k_BT <<1$ and $\sim \mathrm{ln} (g\mu_B B/k_BT)/1.3$ for $g\mu_B B/k_BT >> 1$. For $T$ = 100 mK and assuming $g$ = 2, this field scale is $B$ = 500 mT. 

The first correction to the conductivity, although independent of $B$, leads to a negative contribution to the resistivity, which we can obtain by inverting the conductivity tensor, and noting that EEI corrections also lead to a contribution in the Hall coefficient, which are twice the corrections to the resistivity due to EEI effects. These corrections, calculated by Houghton \cite{Houghton}, are given as 

\begin{equation}
\Delta R_{EEI}^{ex} (T,B) =  \frac{-m^*}{4 \pi^2 \hbar n \tau} \Big( 2-\frac{3F}{2} \Big)[1-(\omega_c\tau)^2] \mathrm{ln} \Big( \frac{k_BT \tau}{\hbar} \Big)
\end{equation}

As we noted earlier, $\omega_c \tau$ in our high mobility sample, especially at large positive values of $V_g$, can be substantial even at small values of $B$. This discussion makes is clear that for analyzing low field data, we must consider the effect of EEI along with localization. Usually, the procedure is to isolate the EEI contributions by considering large fields, at which localization corrections are negligible. However, for the high mobility STO-based 2DCGs, the classical contribution rapidly increases with increasing field, making the resolution of EEI contributions in this manner impossible. Another way to isolate EEI is by measuring MR in fields parallel to the 2DCG, since this would eliminate the large positive background of $\Delta R_{cl} (B)$. However, in the case of STO-based 2DCGs, this runs into difficulties as one still has to contend with a negative quadratic background from magnetic scattering. It is also possible in principle to isolate the EEI contribution using $R$ vs $T$ data, in cases where the $\Delta R_{EEI}^{ex} (T,B)$ is negligible due to $\omega_c \tau$ being very small. Since EEI leads to a logarithmic increase in $R$ as temperature is lowered, similar to weak localization effects, this is usually done by measuring $R$ vs. $T$ in the presence of a magnetic field larger than that required to suppress localization effects. However, the application of a magnetic field would also affect the $\Delta R_{mag} (T,B)$ contribution in STO-based 2DCGs, making the isolation of EEI effects difficult. We do expect EEI effects in STO-based 2DCGs to be substantial especially at the low temperatures of study, given that the carriers originate from the narrow 3d $t_{2g}$ orbitals of Ti.\\

\noindent{\textit{Contributions due to superconducting fluctuations} ($\Delta R_{SC} (T,B)$): Finally, we discuss the contribution due to superconducting fluctuations. STO-based 2DCGs commonly show a superconducting transition below about 300 mK. In the vicinity of a superconducting transition, Aslamazov-Larkin\cite{Aslamazov} corrections to the conductivity, which are caused by fluctuating Cooper pairs, and Maki-Thompson\cite{Maki} corrections which are caused by the coherent scattering of carriers off the fluctuating Cooper pairs, can be important, depending on sample cleanliness and measurement temperature.\cite{Tinkham} This transition was not observed in the case of our (111) LSAT/STO sample. Our sample does show a very slight drop in resistance below about 300 mK for $V_g \geq$ 60 V (see Fig. 1), but it is not clear whether this slight drop is due to superconducting fluctuations, or due to other contributions to the resistivity, such as antilocalization corrections or magnetic scattering. Due to the absence of a full superconducting transition, we ignore this effect in our analysis, however, we note that it must be taken into consideration in samples which do show a superconducting transition.}

\section{Sample fabrication and measurement}
Pulsed laser deposition was used to deposit 12 monolayers of LSAT epitaxially on (111) oriented STO at a partial oxygen pressure of 10$^{-4}$ Torr.\cite{Huang} No post growth annealing step was performed. Using a combination of photolithography and Ar ion milling, the 5 mm $\times$ 5 mm LSAT/STO chip was patterned to make four Hall bars, 100 $\mu$m wide and 600 $\mu$m long. Two of the Hall had their lengths oriented along the [1$\bar{1}$0] crystal direction and the other two had their lengths oriented along the [$\bar{1}\bar{1}$2] crystal direction. Ti/Au was deposited on contact pads, and Al wirebonds were made to allow for a 4-probe measurement configuration of transverse and longitudinal resistance. The sample was attached to a copper puck using silver paint, with care being taken to keep the silver paint off the sides of the sample, which enabled the application of a back gate voltage. The sample was measured in an Oxford Kelvinox MX100 dilution refrigerator. Standard lockin measurement techniques were used to measure the differential resistance, with an ac frequency of 3 Hz, and an ac current $\sim$ 100 nA. We have shown in an earlier publication\cite{Bal2} that transport in (111) LSAT/STO samples grown under these conditions does not exhibit the directional anisotropy which characterizes transport in (111) LAO/STO.\cite{Davis} Hence we only discuss data obtained on a single Hall bar, oriented along the [$\bar{1}\bar{1}$2] direction, other Hall bars showing qualitatively similar results. On initially cooling down to $T$ = 50 mK, $V_g$ was swept multiple times over the entire range, 100 V to -40 V, in order to ensure that the changes in properties due to changes in $V_g$ are reproducible, going always from higher to lower values of $V_g$. 


As discussed earlier, $R$ as a function of $T$ can be nonmonotonic for STO-based 2DCGs. The low $T$ dependence for our sample is as shown in Fig. 1. The data do show low $T$ increase and saturation of resistance, but full superconductivity is not observed even at the most positive values of $V_g$ used. We study the MR below $T$ = 750 mK for $V_g \geq$ 60 V, since in this range of $V_g$, the hysteretic MR associated with a ferromagnetic phase is absent,\cite{Bal} hence allowing the analysis of the low field MR in terms of weak localization.

\begin{figure}[ht]
      \begin{center}
      \includegraphics[width=0.48\textwidth]{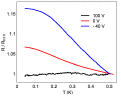}
      \caption{$R$ as a function of $T$ for $V_g$ = 100 V, 0 V, and -40 V. $R$ has been normalized to its value at $T$ = 500 mK.}
      \label{fig1}
     \end{center}
\end{figure}

\section{Magnetoresistance analysis}
At millikelvin temperatures, the $\Delta R_{ph} (T)$ and $\Delta R_{ion} (T)$ contributions freeze out, hence they can be ignored in our analysis. $\Delta R_{SC} (T,B)$ is also ignored since we do not see superconductivity in our sample, leaving us with the following equation for $R(T,B)$:
\begin{multline}
R(T,B) = R_0 + \Delta R_{cl} (T,B) + \Delta R_{mag} (T,B) \\
+ \Delta R_{EEI} (T,B) + \Delta R_{loc} (T,B)
\end{multline}
For $V_g \geq$ 60 V, as discussed earlier, we can approximate $R$ as $R_0$, and hence write the differential MR as:
\begin{equation}
\frac{\delta R(T,B)}{R} = \frac{R(T,B) - R(T,B=0)}{R},
\end{equation}
which has the terms $\Delta R_{loc} (T,B) - \Delta R_{loc} (T,B=0)$. From Eqn. 2, and noting that the asymptotic form for $\Psi (1/2 + B_\alpha/B)$ as ${B \to 0}$ is $\mathrm{ln} (B_\alpha/B)$, one arrives at the following form for the differential localization correction:\cite{Santhanam}

\begin{equation}
\frac{\delta R_{loc} (T,B)}{R} = -\frac{3}{2} f (B,B_2) +
\frac{1}{2} f(B,B_1). 
\end{equation}

Here the first term is the triplet Cooperon contribution while the second term is the singlet Cooperon contribution, and the function $f$ is given as:

\begin{equation}
f(B,B_\alpha) = \frac{R}{2 \pi^2 \hbar/e^2} \Big[\Psi \Big( \frac{1}{2} + \frac{B_\alpha}{B} \Big) - \mathrm{ln} \Big( \frac{B_\alpha}{B} \Big) \Big].
\end{equation}

Here we note that the elastic field $B_0 = \hbar / 4eD\tau$ does not feature in the above equations, thus removing the dependence on $\tau$, which is difficult to determine, as we discussed earlier. Figure 2 shows $\delta R/R$ for $V_g$ = 70 V and $T$ = 50 mK, along with a fit to Eqn. 7 and 8. We see that attempting to fit the low field increase in the MR, which is associated with the presence of a strong SOI, leads to an extremely poor fit at higher fields. This excess negative MR cannot be explained by the classical quadratic background alone since that gives a positive MR. Hence, in our analysis, we use the following terms to account for the background due to the classical MR, magnetic scattering, as well as EEI:

\begin{equation}
\frac{\delta R_{BG} (T,B)}{R} = AB^2 - \frac{CB^2}{D + EB^2}.
\end{equation}

\begin{figure}[ht]
      \begin{center}
      \includegraphics[width=0.48\textwidth]{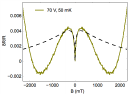}
      \caption{$\delta R/R$ vs $B$ at $V_g$ = 70 V and $T$ = 50 mK. The dashed line is a fit to Eqn. 2.}
      \label{fig2}
     \end{center}
\end{figure}

\begin{figure}
      \begin{center}
      \includegraphics[width=0.48\textwidth]{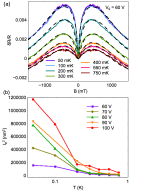}
      \caption{\textbf{a} $\delta R/R$ vs $B$ at $V_g$ = 60 V and various $T$. The dashed line is a fit to Eqn. 3. with a background described by Eqn. 7 \textbf{b} $l_\phi^2$ ($\sim \tau_\phi$) as a function of $T$ on a logarithmic scale, for various $V_g$.}
      \label{fig3}
     \end{center}
\end{figure}

\begin{figure}
      \begin{center}
      \includegraphics[width=0.46\textwidth]{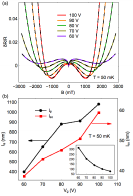}
      \caption{\textbf{a} $\delta R/R$ vs. $B$ at $T$ = 50 mK and various values of $V_g$. The dashed line is a fit to Eqn. 3. with a background described by Eqn. 7. \textbf{b} $l_\phi$ and $l_{so}$ as a function of $V_g$ for $T$ = 50 mK. Inset shows the variation of the sheet resistance with $V_g$ at $T$ = 50 mK.}
      \label{fig4}
     \end{center}
\end{figure}

The positive classical orbital background, and the typically negative EEI contribution due to exchange and singlet Hartree terms described by Eqn. 5, are quadratic in field, and are accounted for by the first term of Eqn. 9. Since this term is proportional to $(\omega_c \tau)^2$, we expect it to be stronger for larger values of $V_g$. We fixed the value of the coefficient $A$ for all temperatures for a particular $V_g$. Although the classical contribution can be $T$-dependent due to changes in $\tau$, and the exchange and the singlet Hartree contributions of EEI, vary logarithmically with $T$, we found experimentally that our background was well described by assuming a temperature independent parameter $A$. This leads us to conclude that the effective temperature dependence of $\tau$ and this particular EEI contribution is likely to be small in this temperature range. 

The second term in Eqn. 9 is quadratic at small fields and saturates at higher fields. This term might arise from the negative contribution due to magnetic scattering discussed earlier. Contributions to EEI which come from the triplet exchange interactions, described by the second term of Eqn. 4, and which can be positive or negative, are quadratic at smaller fields, and logarithmic at larger fields. This can also be roughly approximated by the second term of Eqn. 9. The coefficients $C$, $D$, and $E$ were allowed to vary with $T$, since the contributions due to magnetic scattering and EEI can be temperature dependent.

$B_{so}$ was also held constant since it is expected to be independent of $T$, given that the factors contributing to SOI, namely, band structure effects, atomic SOI, and applied electric field due to $V_g$ are expected to be constant in this temperature range.  Zeeman and magnetic scattering effects on weak localization were ignored. 

Figure 3a shows the MR data, at various values of $T$ for $V_g$ = 60 V, with fits to Eqn. 2, along with a background given by Eqn. 6. Figure 3b shows the variation of the extracted $l_\phi^2 \sim \tau_\phi$ as a function of $T$, for various values of $V_g$. We see that $\tau_\phi$ increases as $T$ is decreased for all $V_g$, and as expected, the increase is the largest for $V_g$ = 100 V for which sheet resistance $R$ is the smallest. $\tau_\phi$ also seems to saturate at lower values of $T$, with the biggest effect for $V_g$ = 60 V, which has the largest $R$. 

Figure 4a shows the MR for various values of $V_g$, measured at $T$ = 50 mK. From these fits we obtained estimates for $l_{so}$ and $l_\phi$ at $T$ = 50 mK for values of $V_g$, which are plotted in Fig. 4b. We see that $l_\phi$, on the left axis, decreases as a function of $V_g$. This is as expected from the variation in sheet resistance with $V_g$, shown in the inset of Fig. 4b. However, $l_{so}$ clearly also decreases with $V_g$, indicating that SOI becomes stronger as $V_g$ is reduced. Further decrease in $V_g$ at these temperatures leads into a ferromagnetic phase characterized by a hysteretic MR, as shown in our earlier report.\cite{Bal} 

At first glance, this variation in SOI may seem counter-intuitive if the SOI is Rashba type, with the Rashba Hamiltonian given as $H_R = \alpha(\hat{\textbf{n}} \times \vec{\textbf{k}}) \cdot \vec{\textbf{S}}$. Here $\vec{\textbf{S}}$ are the Pauli matrices, $\vec{\textbf{k}}$ is the electron wave vector, and $\hat{\textbf{n}}$ is the unit vector perpendicular to the 2DCG plane. The Rashba SOI coupling constant $\alpha$ is dependent on $V_g$, hence one would expect SOI to increase with $V_g$.\cite{Caviglia, Miller} However, one may explain the observed trend by noting the effect of atomic SOI. Density functional theory calculations have demonstrated that in the presence of the trigonal crystal field experienced by the Ti ions near the interface of (111) oriented STO-based systems, the three 3$d$ $t_{2g}$ orbitals split into an $e_g'$ doublet and an $a_{1g}$ singlet.\cite{Pickett,Xiao} The ordering of these orbitals is determined by strain. It is also known that the effects of atomic SOI are strongest for degenerate orbitals, leading to their mixing and splitting.\cite{Joshua} In the (111) LSAT/STO sample studied, if the $e_g'$ doublet is lower in energy than the $a_{1g}$ singlet, this might explain the observed increase of SOI at lower carrier concentrations, when the low energy bands are preferentially filled.

\section{Conclusions}
We have shown that analysis of MR in STO-based 2DCGs must consider a variety of scattering phenomena which have complicated field and temperature dependence. The strong negative MR shown by our LSAT/STO sample suggests that magnetic scattering and EEI contributions must play a major role in determining MR. Analysing our data in terms of these contributions in addition to a positive classical background and weak localization effects, leads us to conclude that SOI indeed gets stronger at smaller gate voltages, and may play a role in the ferromagnetic state that develops at these gate voltages. We note that we have neglected the contribution of magnetic scattering and Zeeman effect to the weak localization corrections. We have also neglected the fact that converting from resistance to conductance involves considering the Hall angle, which can lead to a 20\% difference in the estimated value of conductance, and can affect the weak localization contribution.

The U.S. Department of Energy, Office of Basic Energy Sciences supported the work at Northwestern University through Grant No. DE-FG02-06ER46346. Work at NUS was supported by the MOE Tier 1 (Grants No. R-144-000-364-112 and No.R-144-000-346-112) and Singapore National Research Foundation (NRF) under the Competitive Research Programs (CRP Awards No. NRF-CRP8-2011-06, No. NRF-CRP10-2012-02, and No. NRF-CRP15-2015-01).  This work utilized Northwestern University Micro/Nano Fabrication Facility (NUFAB), which is supported by the State of Illinois and Northwestern University.

$^*$v-chandrasekhar@northwestern.edu

\end{document}